\documentstyle[12pt]{article}
\input psbox
\newcommand{\beq}{\begin{equation}}
\newcommand{\eeq}{\end{equation}}
\newcommand{\beqn}{\begin{eqnarray}}
\newcommand{\eeqn}{\end{eqnarray}}
\begin{document}
\thispagestyle{empty}

\begin{flushright}
{\parbox{3.5cm}{
UAB-FT-372

August, 1995

hep-ph/9508339
}}
\end{flushright}

\vspace{3cm}
\begin{center}
\begin{large}
\begin{bf}
THE TOP-QUARK WIDTH IN THE LIGHT OF $Z$-BOSON PHYSICS
\footnote{Based on the talk presented at the
{\it Workshop on Physics of the Top Quark}, Iowa State University,
 May 1995.}\\
\end{bf}
\end{large}
\vspace{1cm}
Joan SOL\`A\\

\vspace{0.25cm}
Grup de F\'{\i}sica Te\`orica\\
and\\
Institut de F\'\i sica d'Altes Energies\\
\vspace{0.25cm}
Universitat Aut\`onoma de Barcelona\\
08193 Bellaterra (Barcelona), Catalonia, Spain\\
\end{center}
\vspace{0.3cm}
\hyphenation{super-symme-tric}
\hyphenation{com-pe-ti-ti-ve}
\begin{center}
{\bf ABSTRACT}
\end{center}
\begin{quotation}
\noindent
We discuss possible non-standard contributions to the top-quark width,
particularly the virtual effects on the standard
decay $t\rightarrow W^+\,b$ within the context of the MSSM.
We also place a renewed
emphasis on the unconventional mode $t\rightarrow H^+\,b$ in the light of
recent analyses of $Z$-boson observables. It turns out that in the region of
parameter space highlighted by $Z$-boson physics, the charged Higgs mode
should exhibite an appreciable branching fraction as
compared to the standard decay of the top quark.
Remarkably enough, the corresponding quantum effects
in this region are also rather large, slowly decoupling, and most likely
resolvable in the next generation of experiments at Tevatron and at LHC.

\end{quotation}

\newpage

\baselineskip=6.5mm

The recent discovery of a heavy top quark ($m_t\sim 180GeV$)
at Tevatron\,\cite{Tevatron} constitutes, paradoxically as it may sound,
both a reassuring confirmation
of a long- standing prediction of the Standard Model (SM) of the
electroweak interactions and, with no less emphasis, the consolidation
of an old and intriguing suspicion, namely,
that the SM cannot be the last word in elementary particle physics.
Needless to say, the ultimate proof of this conjecture can only be
substantiated in the experimental ring.

In the meanwhile, the projected ten-fold increase of the Tevatron
luminosity via the Main Injector and
Recycler facilities in combination with a $\sim 40\%$
increase of the top-quark production cross-section at a $2\,TeV$ running
energy, as compared to the $1.8\,TeV$ run (Run I),
augur a brilliant Run II performance of this machine; perhaps also an exciting
new era of top-quark physics that may explicitly reveal the long
sought-after signs of newness and non-standardness.
In fact, the finding of such a heavy quark poses some questions of
fundamental nature that go beyond the SM,
may be the most obvious one being the following:
is the top quark an ``abnormal'' fermion or on the contrary it is
the only ``normal'' fermion in the SM?. Supporters of the first contention
may argue that the top quark is the only superheavy quark in the SM, whereas
opponents may adduce that the top quark is the only quark whose Yukawa coupling
with the SM Higgs boson is of the same order as the electroweak gauge coupling.
Be as it may, this same question formulated in a non-SM context may result in a
richer panoply of answers. For instance, in topcolour
models\,\cite{Hill1} the huge mass of the top quark is linked to the
postulation of a new strong force responsible for the formation of a dynamical
quark mass; thereupon it is
not inconceivable to think of the
spontaneous symmetry breaking mechanism (SSB) in the SM as a phenomenon of
top-quark condensation.
In more sophisticated versions, like in the so-called
topcolour-assisted models\,\cite{Hill2},
whether technicolour\,\cite{Technicolour}-like
or Higgs-like, one relaxes the requirement that a
top condensate accounts for the full SSB and makes allowance for a hierarchy
of electroweak mass scales. Thus,
on the one hand, the large mass $m_t$ is almost entirely
driven by a new (non-confining) strong interaction, preferentially coupled
to the third quark generation, which gives rise to a dynamical condensate
naturally tilted in the top-quark direction; light
fermion masses, on the other hand, are generated by the underlying
(extended\,\cite{ETC}, perhaps even walking\,\cite{walking})
technicolour or by fundamental Higgs interactions.

Quite in contrast, in supersymmetric theories, like the Minimal
Supersymmetric Standard
Model (MSSM)\,\cite{MSSM}, one sticks altogether to fundamental scalars.
However, the corresponding spectrum of higgses and of
Yukawa couplings is far and away richer than in the SM;
and, in such a framework, the
bottom-quark Yukawa coupling may counterbalance the smallness of the bottom
mass at the expense of a large value of $\tan\beta$ --the ratio of the vacuum
expectation values (VEV's) of the two Higgs doublets-- the upshot being that
the top-quark and bottom-quark Yukawa couplings standing in the
superpotential\,\cite{MSSM},
\begin{equation}
h_t={g\,m_t\over \sqrt{2}\,M_W\,\sin{\beta}}\;\;\;\;\;,
\;\;\;\;\; h_b={g\,m_b\over \sqrt{2}\,M_W\,\cos{\beta}}\,,
\label{eq:Yukawas}
\end{equation}
can be of the same order of magnitude, perhaps even showing up
in ``inverse hierarchy'': $h_t<h_b$ for $\tan\beta> m_t/m_b$.
Notice that due to the perturbative bound
$\tan\beta\stackrel{\scriptstyle <}{{ }_{\sim}}60-70$ one never reaches
a situation where $h_t<<h_b$.
In a sense, $h_t\sim h_b$  could be judged as a natural
relation in the MSSM, and thus some of the criticism raised above loses
its meaning since $g\sim h_t\sim h_b$ can be made to
coexist with $m_t>>m_b$ in the MSSM even at the electroweak scale.
As a matter of fact one ends up by rephrasing the same problem by posing a
different question; one no longer asks about why $m_t$ is much larger
than $m_b$, though one has to cope with a rather
large value of $\tan\beta$ that must be explained --may be by
invoking some unification model that subsumes the general
structure of the MSSM.

On the phenomenological side, one should not dismiss the possibility that
the bottom-quark Yukawa coupling could play a
momentous role in the physics of the top quark, to the extend of
drastically changing standard expectations on top-quark observables,
particularly on the top-quark width. In this respect
it should be mentioned that the present measurements
of the branching fraction of the standard decay of the top quark
\cite{branching},
\beq
{\Gamma (t\rightarrow W^+\,b)\over \Gamma (t\rightarrow {\rm all})}
= 0.87^{+0.13+0.13}_{-0.30-0.11}\,,
\eeq
do still leave room enough to accomodate non-standard decays; hence we
may expect additional decay modes of the top quark into bottom jets plus
a new charged pseudoscalar particle subsequently
disintegrating into fermion pairs.
A non-standard mode that has been suggested along this line in the framework of
Refs.\,\cite{Hill1,Hill2} is $t\rightarrow \tilde{\pi}^+\,b$,
where $\tilde{\pi}^+$ is a charged member of the ``top-pion'' triplet,
one of the firmest predictions of topcolour models\footnote{Top-pions are
predicted to be in the top-quark mass range, so it is not clear which one
of the decay modes $t\rightarrow\tilde{\pi}^+\,b$ or
$\tilde{\pi}^+\rightarrow t\,\bar{b}$ is kinematically allowed. One should
be open to both possibilities and be prepared to distinguish them from
the corresponding charged Higgs modes\,\cite{RAJ}.}.
However, in this
case the coupling strength is governed by a Goldberger-Trieman type relation:
$g_{tb\tilde{\pi}}\sim m_t/\sqrt{2}f_{\tilde{\pi}}\sim 2.5$ ( $f_{\tilde{\pi}}$
being the top-pion decay constant), and as a consequence top-pion physics
is basically sensitive
to the top-quark mass, not to the bottom-quark mass.

In this talk we wish to emphasize the possibility that the
charged pseudoscalar involved in a potential unconventional
top-quark decay be the charged Higgs of the MSSM:
$t\rightarrow H^+\,b$\, \footnote{In the MSSM there are
several additional $2$-body decays\,\cite{Hamburg,Japonesos}
of the top quark and also a host of exotic
$3$-body final states worth studying\,\cite{Guasch1}.}. In contrast to
$\tilde{\pi}^+$, the charged Higgs can be, as noted
above, very sensitive to bottom-quark interactions. Specifically,
after expressing the two-doublet Higgs fields of the MSSM in terms
of the corresponding mass-eigenstates,
the interaction Lagrangian describing the $t\,b\,H^{\pm}$-vertex
reads as follows\,\cite{Hunter}:
\beq
{\cal L}_{Hbt}={g\,V_{tb}\over\sqrt{2}M_W}\,H^-\,\bar{b}\,
[m_t\cot\beta\,P_R + m_b\tan\beta\,P_L]\,t+{\rm h.c.}\,.
\label{eq:LtbH}
\eeq
Similarly, from the $D$-type terms of the MSSM Lagrangian
the relevant interaction vertices involving the charged Higgs
and the stop and sbottom squarks take on the form
\beq
{\cal L}_{H\tilde{b}\tilde{t}}=-{g\over \sqrt{2}\,M_W}\,H^-\,\left(
g_{LL}\,\tilde{b}_L^*\,\tilde{t}_L+g_{RR}\,\tilde{b}_R^*\,\tilde{t}_R
+g_{LR}\,\tilde{b}_R^*\,\tilde{t}_L+g_{RL}\,\tilde{b}_L^*\,\tilde{t}_R\right)\
 +{\rm h.c.}\,,
\label{eq:Htildebt}
\eeq
with
\beqn
g_{LL}&=&M_W^2\sin 2\beta -(m_t^2\cot\beta+m_b^2\tan\beta)\,,\nonumber\\
g_{RR}&=&-m_tm_b(\tan\beta+\cot\beta)\,,\nonumber\\
g_{LR}&=&-m_b(\mu+A_b\tan\beta)\,,\nonumber\\
g_{RL}&=&-m_t(\mu+A_t\cot\beta)\,,
\eeqn
$A_{t,b}$ being the trilinear soft SUSY-breaking parameters\,\cite{MSSM}.
Notice that
$\tilde{q'}_a=\{\tilde{q}_L,\tilde{q}_R\}$ are the weak-eigenstate squarks
associated to the two chiral fermion components
 $P_{L,R}\,q\equiv \frac{1}{2}\,(1\mp\gamma_5)\,q$;
they are related to
the corresponding mass-eigenstates $\tilde{q}_a=\{\tilde{q}_1,\tilde{q}_2\}$
by a rotation $2\times 2$ matrix (we neglect intergenerational mixing):
\begin{eqnarray}
\tilde{q'}_a&=&\sum_{b} R_{ab}^{(q)}\tilde{q}_b,\nonumber\\
R^{(q)}& =&\left(\begin{array}{cc}
\cos{\theta_q}  &  \sin{\theta_q} \\
-\sin{\theta_q} & \cos{\theta_q}
\end{array} \right)\;\;\;\;\;\;
(q=t, b)\,.
\label{eq:rotation}
\end{eqnarray}
{}From these Lagrangians it is clear that the parameter $\tan\beta$ is called
to play a fundamental role in Higgs-top-bottom interactions and,
of course, in any SUSY-like version of them\footnote{Notice that the SUSY
interactions specified in the Lagrangian (\ref{eq:Htildebt}) enter
the dynamics of $t\rightarrow H^+\,b$ through
virtual loop corrections (Cf. Fig.1 below).}.
In the MSSM we have a full plethora of additional Higgs-like interactions
originally involving the same
Yukawa couplings (\ref{eq:Yukawas}); namely, the interactions
of matter fermions and sfermions with
higgsinos, the spin $1/2$ (super-) companions of higgses. Because of SSB,
the higgsinos mix with the gauginos (the fermion partners of the gauge bosons)
and in the mass-eigenstate basis
form the so-called charginos and neutralinos. In the end one obtains a
set of supersymmetric vertices of the type
fermion-sfermion-chargino/neutralino involving a fairly
complicated admixture of top and bottom Yukawa couplings
\footnote{A detailed interaction Lagrangian is given
e.g. in eqs.(18)-(19) of Ref.\cite{GJSH}.} affecting the loop structure
of both the conventional decay $t\rightarrow W^+\,b$ and
the unconventional mode $t\rightarrow H^+\,b$.
On the face ot it, it is patent that a $\tan\beta$-enhanced
bottom-quark Yukawa coupling , i.e.
$\tan\beta\stackrel{\scriptstyle >}{{ }_{\sim}}m_t/m_b$, may have a
drastic impact on the MSSM phenomenology of the top-quark.

In spite of the already abundant literature on the various aspects, whether
theoretical or experimental, of the charged Higgs decay of
the top quark\,\cite{Hamburg,Hunter,tbHexp},
we believe that nowadays it definitely deserves a renewed interest.
There are in part new theoretical reasons, but above all there are intriguing
phenomenological motivations grounded on the most recent results from
the high precision world of $Z$-boson physics.
Theoretically, the large $\tan\beta$ regime is naturally suggested
in top-bottom-tau Yukawa coupling unification models as well as in
many string-like unification schemes where, in order to get the radiative
electroweak symmetry-breaking pattern, one is forced to
depart from the canonical universal boundary conditions on the scalar masses
at the GUT scale\,\cite{string}.
Most remarkable, the large $\tan\beta$ regime in conjunction with a moderate
value of the CP-odd Higgs mass around $m_{A^0}\simeq 50\,GeV$
has been insistently
projected by phenomenological analyses of $Z$-boson observables
within the context of the MSSM, such as in comprehensive global fits
of electroweak precision data\,\cite{Chankowski}
and in thorough scrutinies of the MSSM parameter
space\,\cite{Rb1}-\cite{Rb2}.
These studies were aimed at solving, or at least alleviating, the
discrepancies (at the $2-3\,\sigma$ level) between the strict SM theoretical
prediction and the corresponding experimental measurements of several
$Z$-boson observables,
most conspicuously the $R_b$, $R_c$ branching ratios
and the lineshape value of the strong coupling constant at the scale of the
$Z$-boson
mass, $\alpha_s(M_Z)$; and they have prompted some speculations on
new physics\,\cite{ErlerLang}, e.g. claiming the existence
of relatively light sparticles\footnote{For
alternative --topcolour and related-- approaches to the
$R_b$ anomaly, see e.g. Ref.\cite{Zhang}.}
\,\cite{Shifman,GS2}\footnote{Although the analysis
of Ref.\cite{Kaneboys} also emphasizes the role played by light sparticles,
it misses -- in contrast to the systematic approach
 of Refs.\cite{GJS1}-\cite{GS1},\cite{GS2}-- crucial quantum effects,
originated in the large $\tan\beta$ region from the neutral components of
the MSSM Higgs sector, which should by no means be understated.}.

In the aforementioned region of parameter space one finds, at the tree-level,
the ratio
\beq
{\Gamma_0 (t\rightarrow H^+\,b)\over \Gamma_0 (t\rightarrow W^+\,b)}=
{\left(1-{M_{H^{+}}^2\over m_t^2}\right)^2\,
\left[{m_b^2\over m_t^2}\,\tan^2\beta+\cot^2\beta\right]
\over
\left(1-{M_W^2\over m_t^2}\right)^2\,\left(1+2{M_W^2\over m_t^2}\right)}\,.
\label{eq:ratioHW}
\eeq
We see from it that if $M_{H^{\pm}}\simeq M_W$ (by the way, a situation
perfectly compatible with $m_{A^0}\simeq 50\,GeV$\,\cite{Hunter}),
there are two regimes of $\tan\beta$ where the
width of the charged Higgs decay becomes of the same order as
(or larger than) the conventional decay width: namely,
i) for $\tan\beta\leq 1$, and
ii) for $\tan\beta\geq m_t/m_b\sim 36$.
The critical status of the decay
$t\rightarrow H^+\,b$ occurs at the intermediate value
$ \tan\beta= \sqrt{m_t/m_b}\sim 6$, where its partial width has
a pronounced dip. Around this value, the charged Higgs mode is
overwhelmed by the canonical mode $t\rightarrow W^+\,b$.
Sufficiently away from the dip, however, $t\rightarrow H^+\,b$
becomes competitive with $t\rightarrow W^+\,b$.
As mentioned above, we do have some theoretical and
experimental motivations \,\cite{GS1},\cite{GS2}
to contend both that  $M_{H^{\pm}}={\cal O}(M_W)$ and that at least
one of the two $\tan\beta$ regimes i) or ii) applies, most probably the latter.

In view of the potential interest of the decay mode $t\rightarrow H^{+}\, b$,
one would naturally like to address the computation of the virtual loop
corrections to its partial width. Of these, the conventional QCD corrections
have already been considered in detail in Ref.\cite{CD} and they turn out
to be  sizeable and negative (of order $-10\%$).
Although they are blind to the
nature of the underlying Higgs model, they need to be subtracted from the
experimentally measured number in order to be be able to
probe the existence of new sources of quantum effects
beyond the SM. These effects may ultimately reveal whether the charged Higgs
emerging from that decay is supersymmetric or not.
Similarly, the SM one-loop corrections to $t\rightarrow W^+\,b$ are
known; they are basically dominated by the QCD gluonic contributions
($\simeq -8\%$)\,\cite{Kuhn} plus small
($\simeq +1\%$) electroweak corrections\,\cite{Denner}.
Thus we may concentrate on just the MSSM {\it additional} loop diagrams.
Here we shall focus on the strong and electroweak SUSY corrections to the
standard top-quark decay $t\rightarrow W^+\,b$\,\,\cite{GJSH},\cite{DHJJS},
and on the strong SUSY
corrections to the unconventional decay\,
$t\rightarrow H^+\,b$\,\cite{Guasch2}.
The direct, process-dependent, SUSY Feynman diagrams contributing to
these decays are sketched in Fig.1.
The analysis of the larger and far more complex
body of SUSY-electroweak Feynman diagrams contributing
to $t\rightarrow H^+\,b$, namely the
corrections mediated by squarks, sleptons, chargino-neutralinos and the
Higgs bosons themselves, is currently under study and
will soon be available\,\cite{CGGJS}.
\begin{figure}

\begin{center}
\newcommand{\vcpsboxto}[3]{{$\vcenter{{\psboxto(#1;#2){#3}}}$}}
\vcpsboxto{12cm}{0cm}{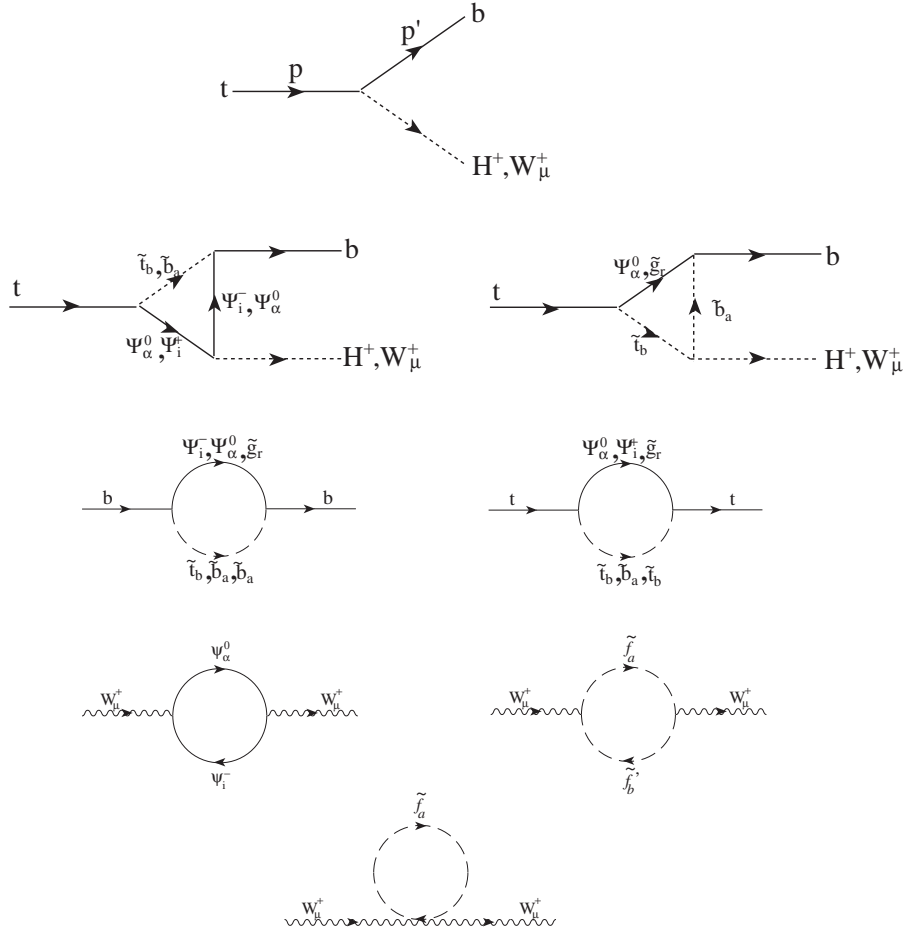}
\end{center}

\vspace{2cm}
\caption
{Feynman diagrams for the one-loop electroweak and strong  SUSY  corrections
to $t\rightarrow W^+\,b$ and $t\rightarrow H^{+}\,b$.
Loop diagrams are summed over all possible
values of the mass-eigenstate charginos
($\Psi^{\pm}_i\,; i=1,2$), neutralinos
($\Psi^{0}_{\alpha}\,; \alpha=1,2,...,4$), stop and sbottom squarks
($\tilde{b}_a, \tilde{t}_b\,; a,b=1,2$) and gluinos
 ($\tilde{g}_r\,; r=1,2,...,8$). Virtual Higgs contributions are
 not depicted\ (Cf. Ref.[33]). }

\end{figure}


The basic free parameters of our analysis, in the electroweak sector, are
contained in the
stop and sbottom mass matrices:
\begin{equation}
{\cal M}_{\tilde{t}}^2 =\left(\begin{array}{cc}
M_{\tilde{t}_L}^2+m_t^2+\cos{2\beta}({1\over 2}-
{2\over 3}\,s_W^2)\,M_Z^2
 &  m_t\, M_{LR}^t\\
m_t\, M_{LR}^t &
M_{\tilde{t}_R}^2+m_t^2+{2\over 3}\,\cos{2\beta}\,s_W^2\,M_Z^2
\end{array} \right)\,,
\label{eq:stopmatrix}
\end{equation}
\begin{equation}
{\cal M}_{\tilde{b}}^2 =\left(\begin{array}{cc}
M_{\tilde{b}_L}^2+m_b^2+\cos{2\beta}(-{1\over 2}+
{1\over 3}\,s_W^2)\,M_Z^2
 &  m_b\, M_{LR}^b\\
m_b\, M_{LR}^b &
M_{\tilde{b}_R}^2+m_b^2-{1\over 3}\,\cos{2\beta}\,s_W^2\,M_Z^2
\end{array} \right)\,,
\label{eq:sbottommatrix}
\end{equation}
These mass matrices are diagonalized by means of the rotation matrices
(\ref{eq:rotation}). We have defined
\beq
M_{LR}^t=A_t-\mu\cot\beta\,, \ \ \ \ M_{LR}^b=A_b-\mu\tan\beta\,,
\eeq
$\mu$ being the SUSY Higgs mass parameter in the
superpotential\footnote{Its
sign is relevant in the numerical analysis. We have corrected a
known inconsistency in this sign as it appears in Ref.\cite{Hunter}, and
have fixed it as in eq.(3) of Ref.\cite{GJSH}.}.
The $M_{{\tilde{q}}_{L,R}}$ are soft SUSY-breaking masses\,\cite{MSSM};
by $SU(2)_L$-gauge invariance we must have $M_{\tilde{t}_L}=M_{\tilde{b}_L}$,
whereas $M_{{\tilde{t}}_R}$, $M_{{\tilde{b}}_R}$ are in general independent
parameters. In the strong supersymmetric sector, the basic parameter is the
gluino mass, $m_{\tilde{g}}$, and the interaction Lagrangian defining the
SUSY-QCD gluino interactions with squarks is the following:
\begin{equation}
{\cal L}= -{g_s\over\sqrt{2}}\,\left[\tilde{q}^{i *}_{L}\,(\lambda_r)_{ij}\,
\bar{\tilde{g^r}}\,P_L\,q^j-\bar{q}^i(\lambda_r)_{ij}\,
P_L\,{\tilde{g^r}}\,\tilde{q}^{j}_{R}\right]+{\rm h.c.}\,,
\end{equation}
where $\tilde{g}^r (r=1,2,...,8)$ are the Majorana gluino fields, and
$(\lambda_r)_{ij} (i,j=1,2,3) $ are the Gell-Mann matrices.

Next we shortly review\footnote{For renormalization niceties,
detailed one-loop formulae and exhaustive numerical
analyses, see Refs.\cite{GJSH},\cite{DHJJS}-\cite{CGGJS}.}
the actual results of our comparative analysis of the SUSY
quantum effects on the partial widts $\Gamma (t\rightarrow W^+\,b)$ and
$\Gamma (t\rightarrow H^+\,b)$
in the on-shell renormalization scheme\,\cite{BSH}.
For convenience
we define the relative correction,
\begin{equation}
\delta={\Gamma-\Gamma_0\over \Gamma_0}\,,
\label{eq:delta}
\end{equation}
with respect to the corresponding tree-level width, $\Gamma_0$.
In Fig.2
$\delta^{SUSY}$ and $\delta_{\tilde{g}}$ refer to the SUSY-electroweak and
SUSY-gluino (i.e. SUSY-QCD) corrections, respectively, both given in the
$G_F$-scheme\,\cite{GJSH}, viz.
parametrized in terms of $G_F$ (Fermi's constant in $\mu$-decay) by using
\beq
{G_F\over\sqrt{2}}={\pi\alpha\over 2 M_W^2 s_W^2}
(1+\Delta r^{MSSM})\,,
\label{eq:DeltaMW}
\eeq
where $\Delta r^{MSSM}$ is the prediction of the $\mu$-decay parameter
$\Delta r$ in the MSSM\footnote{A dedicated study of
$\Delta r^{MSSM}$ has been presented in Ref\,\cite{GS}.}.
Clearly, the term $\Delta r^{MSSM}$ is only relevant for the electroweak
corrections (Fig.2a) and is does not affect the strong contributions
(Figs.2b-2c) due to the absence of one-loop QCD effects in $\mu$-decay.

\begin{figure}
\newcommand{\vcpsboxto}[3]{{$\vcenter{{\psboxto(#1;#2){#3}}}$}}

\begin{tabular}{ll}
\vspace{-4cm}
\hspace{-1.8cm} \vcpsboxto{10cm}{0cm}{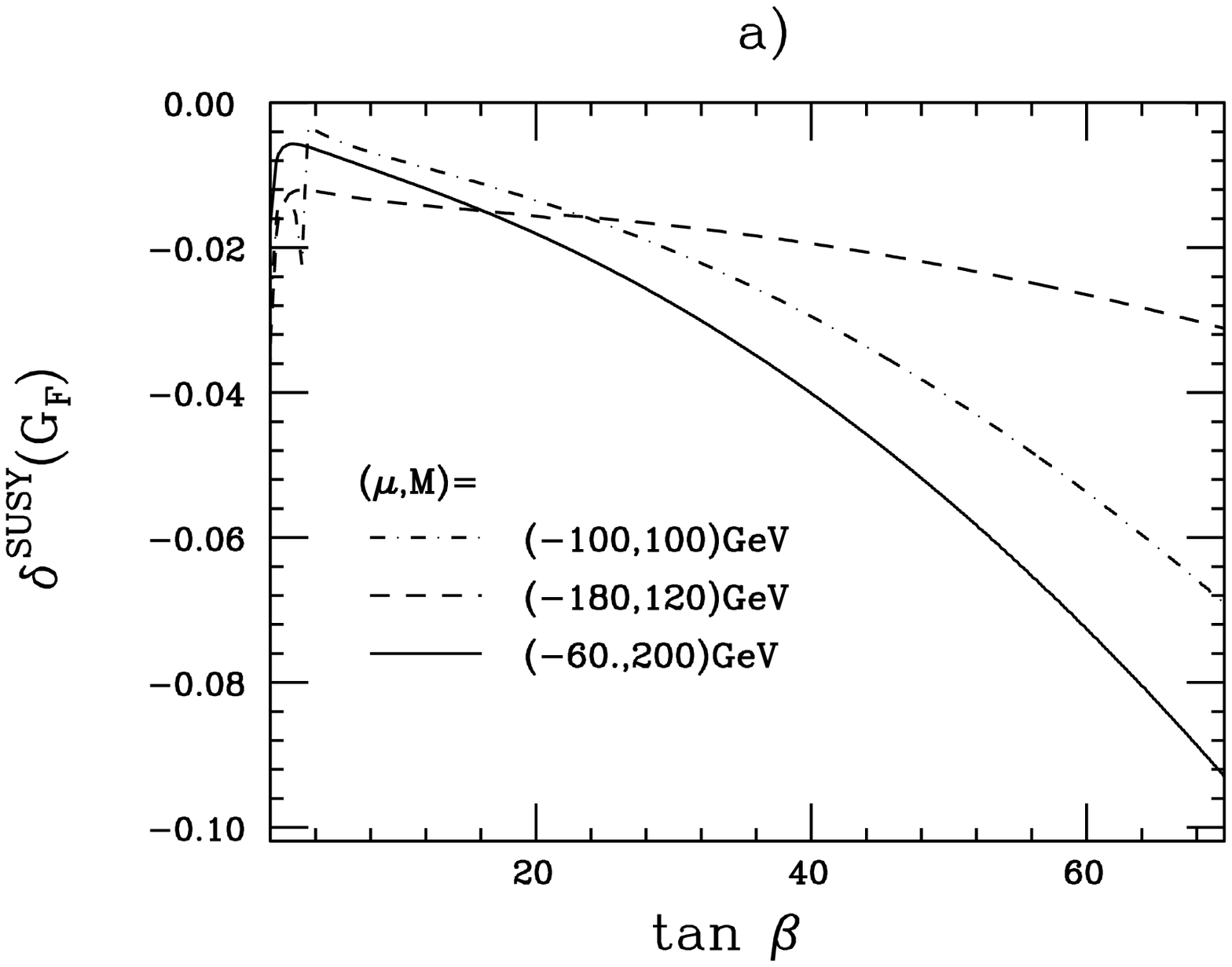} &
\hspace{-1.8cm} \vcpsboxto{10cm}{0cm}{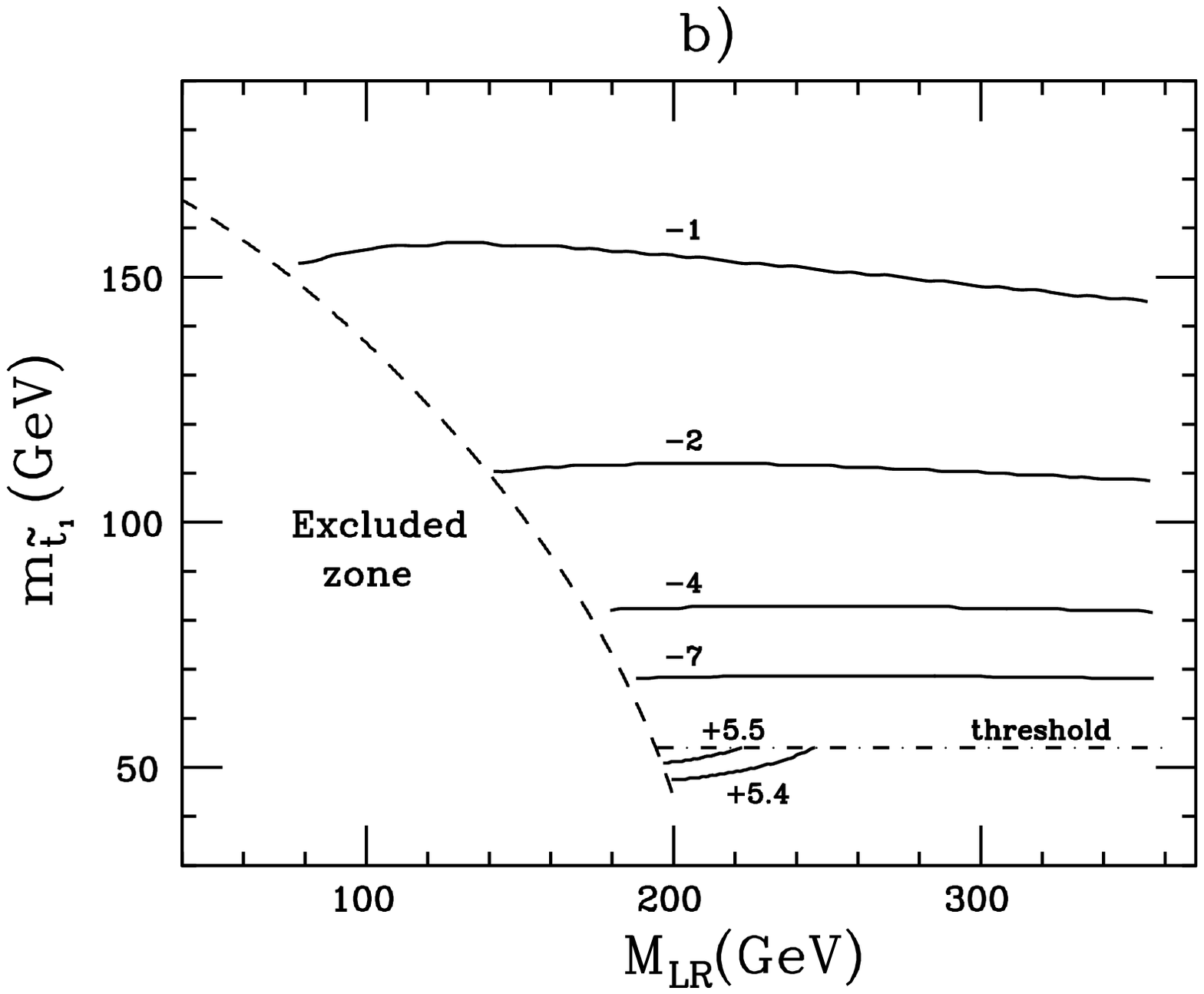} \\
\end{tabular}
\vspace{-1.8cm}
\begin{center}
\vcpsboxto{8cm}{0cm}{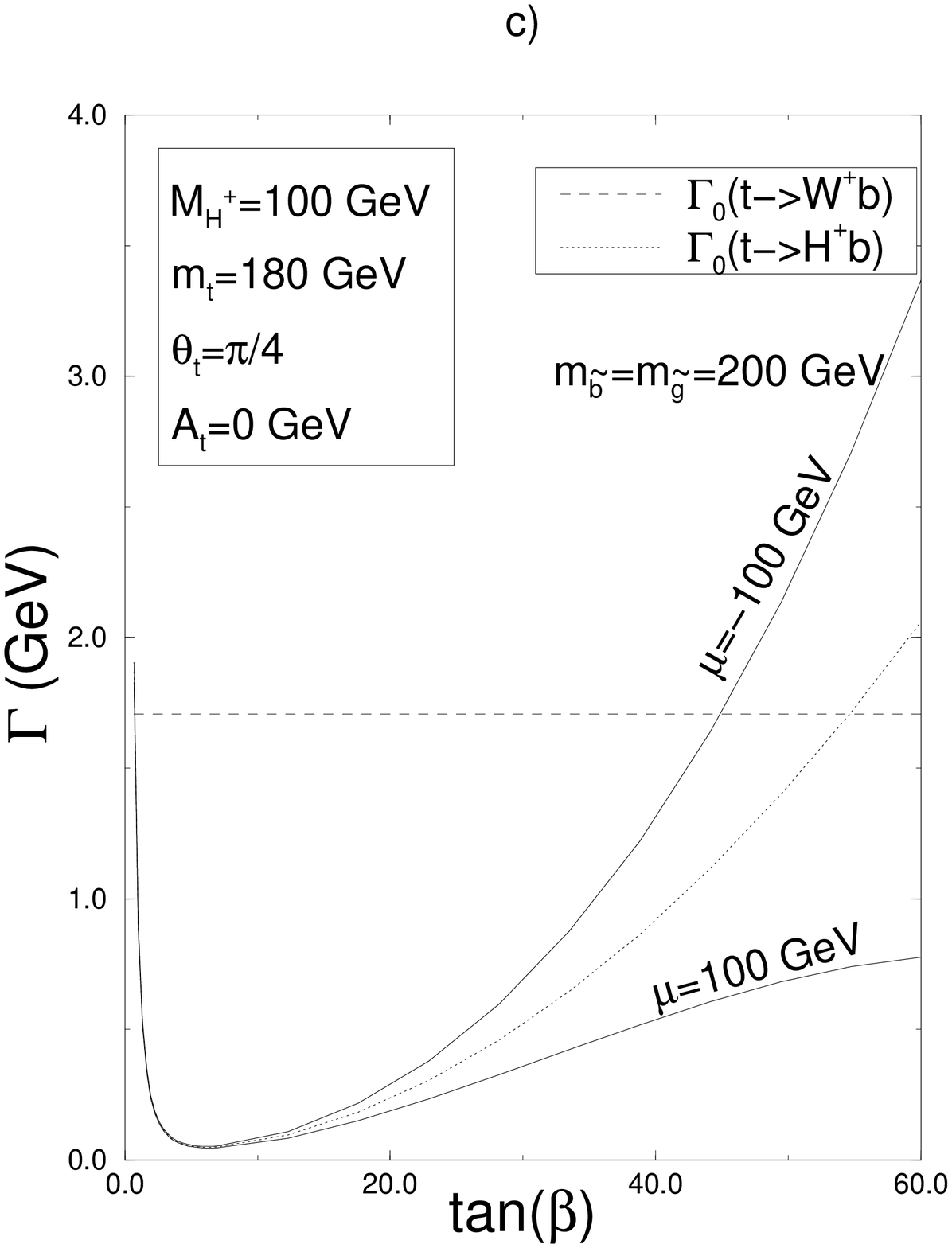}
\end{center}


\caption
{(a) Electroweak SUSY corrections to $\Gamma (t\rightarrow W^+\,b)$ as
a function of $\tan\beta$ for three values of $(\mu,M)$ and fixed sfermion
masses; (b) Isolines in the $(m_{\tilde{t}_1},M_{LR}^t)$-plane
for the SUSY-QCD corrections $\delta_{\tilde{g}}$
(in \%)  to the decay $t\rightarrow W^+\,b$.
(c) SUSY-QCD corrected width of
$t\rightarrow H^+\,b$ versus $\tan\beta$ for $\mu=\pm 100\,GeV$.
Also indicated are the tree-level widths for the standard and charged Higgs
decay modes.}
\end{figure}
A crucial parameter to be tested is $\tan\beta$.
In Fig.2a we plot the SUSY-electroweak corrections to the standard decay
$\Gamma (t\rightarrow W^+\,b)$ as a function of $\tan\beta$ for given choices
of the other parameters\,\cite{GJSH}. We see that they can be of
order $-10\%$ for very large $\tan\beta$. This is in
contrast to the corrections from
the two-doublet Higgs sector of the MSSM where in comparable conditions they
are one order of magnitude smaller\,\cite{HollikHoang}, as it
is also the case with those from the
one-doublet Higgs sector of the SM\,\cite{Denner}.
Noteworthy is also the fact that
the supersymmetric electroweak corrections are of the same
(negative) sign and could
be of the same order of magnitude as the conventional QCD corrections
\,\cite{Kuhn}.
On the whole the standard QCD plus SUSY-electroweak
corrections to $t\rightarrow W^+\,b$
could reduce this partial width
by about $10-15\%$. Consequently, a measurable
reduction beyond $\sim 8\%$ (QCD)
could be attributted to a ``genuine'' SUSY effect.
The fact that the chargino-neutralino sector of the MSSM could afford a
non-negligible quantum correction to the top quark decay width,
in contradistinction to the inappreciable
yield from the scalar Higgs sector of the MSSM, can be traced to the
highly constrained structure of the Higgs potential as dictated
by SUSY\,\cite{Hunter}.

In Fig.2b we study the SUSY-QCD corrections to the standard decay
$t\rightarrow W^+\,b$.
Here we have fixed $m_{\tilde{b}}=m_{\tilde{g}}=120\,GeV$
for the sbottom and gluino masses
and plot contour lines of constant $\delta_{\tilde{g}}$ in
the $(M_{LR}^t,m_{\tilde{t}_1})$-plane\,\cite{DHJJS}.
The excluded zone in Fig.2b violates the condition $M_{\tilde{t}_R}^2>0$
in the stop mass matrix (\ref{eq:stopmatrix}).
Notice that there is a
threshold (pseudo) singularity (dashed line)
associated to the wave-function renormalization of the
top-quark field at $m_{\tilde{t}_1}=54\,GeV$ for $m_t=174\,GeV$,
where $m_{\tilde{t}_1}$ is the lightest stop mass. We cannot arbitrarily
approach this line from above without violating perturbation theory, but
we see that even staying prudentially away from it the SUSY-QCD corrections
are appreciably high ($\stackrel{\scriptstyle <}{{ }_{\sim}}8\%$) and
negative.  On the contrary, if we
approach the threshold line from below (a non-singular limit),
the correction is positive and of order $5\%$.
Nevertheless, it should be clear that the most interesting scenario for our
decay corresponds to $\delta_{\tilde{g}}<0$, since the alternative
two-body supersymmetric decay into stop and gluino,
$t\rightarrow \tilde{{t}_1},\tilde{g}$, is then
phase-space blocked up. This situation is further preferred by the fact that
the strong superymmetric corrections could be reinforced by
the additional negative contributions from the electroweak supersymmetric
sector of the MSSM (Fig.2a). Last but not least, the case
$\delta_{\tilde{g}}<0$ is especial
in that the SUSY-QCD loops would add up to the conventional QCD corrections
($\delta_{QCD}\simeq -8\%$)\,\cite{Kuhn}, so that
in favourable circumstances the total strong
correction could reach $-(15-18)\%$. Therefore,
as the strong SUSY corrections to $t\rightarrow W^+\,b$
are insensitive to $\tan\beta$\,\cite{DHJJS},
we may envision an scenario with
large $\tan\beta$  ($\stackrel{\scriptstyle >}{{ }_{\sim}} m_t/m_b$)
in which the electroweak supersymmetric
corrections, being also negative, are of the same order
of magnitude as the SUSY-QCD contributions
studied here; hence the total MSSM pay-off to the top quark width
--the Higgs correction being negligible\,\cite{HollikHoang}-- could result
in an spectacular reduction
of $\Gamma (t\rightarrow W^{+}\, b)$ by about $25\%$.

In Fig.2c we turn our attention to the alternative decay $t\rightarrow H^+\,b$
and plot the corresponding SUSY-QCD corrected width,
$\Gamma=\Gamma (t\rightarrow H^+\,b)$,
versus $\tan\beta$ for $\mu=+100\,GeV$ and $\mu=-100\,GeV$, and for
given values of the other parameters\,\cite{Guasch2}. In particular,
the stop and sbottom mixing mass
terms are $M_{LR}^t=-\mu\cot\beta$ and $M_{LR}^b=0$, respectively,
and we assume that the two diagonal elements in ${\cal M}_{\tilde{t}}^2$
(resp. in ${\cal M}_{\tilde{b}}^2$ ) are equal.
We see that $\Gamma$
rapidly increases with
$\tan\beta$, the preferred range singled out by the
$Z$ lineshape observables\,\cite{GS1,GS2}.
Highly remarkable is also the incidence of the parameter $\mu$.
Indeed, the sign of  $\delta_{\tilde{g}}$ happens to be opposite
to the sign of $\mu$ and the respective corrections for $\mu$ and for $-\mu$
take on approximately the same absolute value.
The sign dependence of $\delta_{\tilde{g}}$ suggests that
two extreme scenarios could take place with the SUSY-QCD
corrections to $\Gamma (t\rightarrow H^+\,b)$: namely,
they could either significantly enhance the, negative, conventional
QCD corrections\,\cite{CD},
or on the contrary they could counterbalance them and even result in
opposite sign.
Worth noticing is also the dependence of these corrections on the
gluino mass. In Fig.2c we have fixed $m_{\tilde{g}}=200\,GeV$, which is rather
heavy. As a matter of fact, we have checked\,\cite{Guasch2}
that the decoupling rate of the gluinos is
very slow, to the extend that it fakes for a while, so to speak, a
non-decoupling behaviour.
This trait is caused by the presence of a long
sustained local maximum (or minimum, depending on the sign of $\mu$)
spreading over a wide range of heavy gluino masses
centered at $\sim 300\,GeV$\,\cite{Guasch2}. For this reason, heavy gluinos
are in the present instance preferred to light gluinos,
contrary to naive expectations\,\footnote{The existence of
light gluinos of ${\cal O}(1)\,GeV$ is not yet completely excluded.
They could decisively influence the MSSM phenomenology in other instances,
as shown in Ref.\cite{Clavelli}.}.

Let us mention that
we chart significant differences in our analysis as compared to
preliminary calculations in the literature.
In Refs.\cite{LiYangHu} a first study
of the SUSY-QCD corrections to $t\rightarrow H^+\,b$ was presented, but
they neglect a crucial piece of the analysis, viz.
the bottom-quark Yukawa coupling, and as a consequence they are
incorrectly  sensitive to the relevant high $\tan\beta$ corrections
(see Fig.2c). A similar situation
occurs with the incomplete SUSY treatment of
$t\rightarrow W^+\,b$ in Ref.\cite{YangLi}.
Furthermore, the impact
from mixing effects and the incidence of the various parameter dependences
were completely missed and only the simplest
pattern, characterized by degenerate masses,
was considered\footnote{Notice that
the assumption of stop masses equal
to sbottom masses is incompatible with the structure of the mass matrices
(\ref{eq:stopmatrix})-(\ref{eq:sbottommatrix}).}.
Moreover, in the framework of these references, the setting $m_b=0$ makes
the lowest-order width fully proportional to $\cot\beta$; thus, in such a
context, finding quantum effects increasing with $\tan\beta$ is rather useless
since they result in corrections to an uninteresting,
vanishingly small, tree-level width.

In summary, the SUSY contributions to the partial
widths of $t\rightarrow W^+\,b$ and $t\rightarrow H^+\,b$
could be sizeable, especially in the latter decay where they may
comfortably reach several $\pm10\%$ even for ${\cal O}(100)\,GeV$ sparticle
masses. In the former case
an average $\sim 5\%$ (negative) correction seems more realistic; and
although one could also attain the $10\%$ level (and beyond) one has to
wrestle harder with the parameters. Moreover, it should not be understated
the fact that the quantum corrections to these decays hold in a region of the
MSSM
parameter space prompted by the high precision $Z$-boson
observables\,\cite{GS1,GS2},
and in this region $t\rightarrow H^+\,b$
has an appreciable branching ratio as compared to the standard decay
$t\rightarrow W^+\,b$.
The potential size  of the SUSY effects on $t\rightarrow H^+\,b$
stems not only from the strong interaction character of the
SUSY-QCD corrections, but also from the high sensitivity of
this decay mode to the (weak-interaction) SSB
parameter $\tan\beta$. Barring the $\tan\beta<<1$ regime --considered as
very unlikely from the point of view of model building--, we see that
the relevance of the charged Higgs decay mode of the top quark
is ultimately linked to the dynamics of the bottom-quark Yukawa coupling.

All in all, we believe that the Higgs mode
reveals itself as an ideal environment where to study the nature of
the SSB mechanism\footnote{In the event that the charged Higgs boson
is heavier than the top quark, the alternative mode
$H^+\rightarrow t\,\bar{b}$ is expected to exhibit similar
properties\,\cite{RAJ}.}.
We might even venture into saying
that $t\rightarrow H^+\,b$ stands among the best candidate processes
where to target our long and (yet) unsuccessful
search for ``virtual Supersymmetry''; that is, it could be an optimal place
where to enquire for huge, {\it and} slowly decoupling,
quantum supersymmetric effects. In this respect it should be
stressed that the typical size of our corrections is maintained
even for sparticle masses well above the LEP $200$ discovery range.
Theses features are in stark contrast to the standard decay
of the top quark, $t\rightarrow W^+\,b$, whose SUSY-QCD corrections
are largely insensitive to $\tan\beta$ and the corresponding electroweak
corrections are only moderately sensitive to this parameter.
Fortunately,
the next generation of experiments at Tevatron and the future high precision
experiments at LHC may well acquire the ability to test the kind of effects
considered here\,\cite{Atlas}-\cite{CPYuan}. As a very promising example,
we remark the future
measurement of the cross-section for single top-quark production, which is
directly sensitive to the top-quark width\,\cite{CPYuan}.
Thus, in favorable circumstances, we should be able to unravel the existence
of new physics out of a precise measurement of the top-quark width,
or related observables, at a modest --and attainable\,\cite{Fujii,CPYuan}
--precision of $\sim 5-10\%$.
\vspace{0.5cm}

{\bf Acknowledgements}:

\noindent
I thank the organizers for the invitation to
the workshop, where I have had the opportunity to benefit from conversations
with many people, in particular with
P. Chankowski, L. Clavelli, A. Heinson, C. Hill, M. Shifman and X. Zhang.
The excellent logistics by A. Sommerer is also highly appreciated.
I am grateful to my students:  David Garcia and Ricardo Jim\'enez,
for their assistance in the preparation of
this talk; also the concurrence of Toni Coarasa and Jaume Guasch, in
an effort to get ready some preliminary results of Ref.\cite{CGGJS},
is worth mentioning.
This work has been partially supported by CICYT
under project No. AEN93-0474.

\vspace{0.5cm}

\textheight=8.9in

\end{document}